\documentclass[12pt]{article}
\usepackage{cite}
\usepackage{CJK}
\usepackage{amssymb}
\usepackage{amsmath}
\usepackage{enumerate}
\usepackage{graphicx}
\usepackage{amsfonts}
\usepackage{url}
\usepackage{bm}

\usepackage{pifont}
\usepackage{epsfig,subfigure,dsfont,amsthm,amsbsy,mathrsfs,amscd}
\newtheorem{theorem}{Theorem}

\newtheorem{remark}{Remark}

\def\be{\begin{equation}}
\def\ee{\end{equation}}
\def\ba{\begin{array}}
\def\ea{\end{array}}
\input amssym.def

\newcommand\btd{\raise 2pt \hbox{$\hat\bigtriangledown$}\hskip 1.5pt}
\newcommand\bt{\raise 2pt \hbox{$\bigtriangledown$}\hskip 1.5pt}
\begin{document}
\title{\large\bf Entanglement detection via general SIC-POVMs}
\author{Ya Xi$^{1}$ \& Zhu-Jun Zheng$^{1}$ \& Chuan-jie Zhu$^{2}$  \\[10pt]
\footnotesize
\small $^{1}$ Department of Mathematics, South China University of Technology,
GuangZhou 510640,P.R. China\\
\footnotesize
\small $^{2}$ Department of Physics, Renmin Univeisity of China,Beijing, 100872, P.R. China}
\date{}

\maketitle

\bigskip
\begin{abstract}
We study separability problem using general symmetric informationally complete
measurements and propose separability criteria in $\mathbb{C}^{d_{1}}\otimes\mathbb{C}^{d_{2}}$ and $\mathbb{C}^{d_{1}}\otimes\mathbb{C}^{d_{2}}\cdots\otimes\mathbb{C}^{d_{n}}$. Our criteria just require less local measurements and provide experimental implementation in detecting entanglement of unknown quantum states.
\end{abstract}
\bigskip

\medskip
\noindent{\bf 1 Introduction}

\medskip

The detection of entanglement is one of the most fundamental and attractive tasks in quantum information theory and entanglement enables numerous applications ranging from quantum cryptography to quantum computing(see reviews \cite{s1,g09} and the references therein). And there have been some necessary criteria for separability, such as Bell inequality \cite{b64}, positive partial transposition criterion \cite{p96}, realignment criterion \cite{r1,r2,r3}, covariance matrix criterion \cite{c1},and correlation matrix criterion \cite{c2}, entanglement witness \cite{t00}.

Although numerous mathematical tools have been extensively studied, experimental implementation of entanglement detection for unknown quantum states has fewer results \cite{Bell}. The authors \cite{spe12} connected the separability criteria with mutually unbiased bases (MUBs) \cite{s2} in two-qudit, multipartite and continuous-variable quantum systems. Later, Chen \emph{et al.} \cite{s3} proposed separability criteria for arbitrary $d$-dimensional bi-partite states using mutually unbiased measurements (MUMs) \cite{ka14} and  Liu \emph{et al.} \cite{s4} derived separability criteria for multipartite qudit systems,arbitrary high-dimensional bipartite and multipartite systems of multi-level subsystems using sets of MUMs. Another method for the entanglement detection was derived by incomplete sets of MUBs in \cite{ra15}.

Besides mutually unbiased bases, another intriguing topic in quantum information theorey is the symmetric informationally complete positive operator-valued measurements (SIC-POVMs)  \cite{s5}. Most of the literature on SIC-POVMs focus on rank $1$ SIC-POVMs (all the POVM elements are proportional to rank $1$ projectors). Such rank $1$ SIC-POVMs just exist in lower dimensions. The author \cite{s7}introduced the concept of general SIC-POVMs(GSIC-POVMs) in which the elements need not to be of rank one, and Gour and Kalev \cite{gk14} constructed the set of all general SIC-POVMs from the generalized Gell-Mann matrices. In addition, based on entanglement detection,   Chen \cite{clf15} and Shen \cite{shen1} used the GSIC-POVMs to give separability criteria for arbitrary $d$-dimensional bipartite and multipartite systems.

In this paper, we investigate entanglement detection via GSIC-POVMs and propose separability criteria in arbitrary high dimensional bipartite systems of a $d_1$-dimensional subsystem and a $d_2$-dimensional subsystem and multipartite systems of multipartite-level subsystems. The paper is organized as follows. In Section $2$, we recall some basic notions of SIC-POVMs and GSIC-POVMs. In Section $3$, we provide four theorems
based on GSIC-POVMs and formulate the validity and power of entanglement detection. At last, we conclude the paper in Section $4$.

\medskip
\noindent{\bf 2 SIC-POVMs and GSIC-POVMs}

\medskip

A POVM $\{P_{j},j=1,2,\cdots, d^{2}\}$ with $d^2$ rank one operators acting on $\mathbb{C}^{d}$ is symmetric informationally complete, if $P_{j}=\frac{1}{d}|\phi_{j}\rangle\langle\phi_{j}|, j=1,2,\cdots,d^2,$ and $\Sigma_{j=1}^{d^{2}}P_{j}=I,$ where the vectors $|\phi_{j}\rangle$ satisfy $|\langle\phi_{j}|\phi_{k}\rangle|^{2}=\frac{1}{d+1},j\neq k$, and $I$ is the identity operator. The existence of SIC-POVMs in arbitrary dimension d is an open problem. Only in a number of low dimensional cases, the existence of SIC-POVMs has been proved analytically and numerically for all dimensions up to 67(see \cite{s5} and the references therein).

A set of $d^{2}$ positive-semidefinite operators $\{{P}_{\alpha},\alpha=1,2,\cdots,d^{2}\}$ on $\mathbb{C}^{d}$ is said to be a general SIC measurement, if

\begin{eqnarray*}
  (1) \Sigma_{\alpha=1}^{d^{2}}P_{\alpha}=I, \
  (2)\mathrm{Tr}[(P_{\alpha})^{2}]=a,\
  (3)\mathrm{Tr}(P_{\alpha}P_{\beta})=\frac{1-da}{d(d^{2}-1)},
\end{eqnarray*}
where $\alpha,\beta=1,2,\cdots,d^{2}$, $\alpha\neq\beta$, $I$ is the identity operator, and the parameter $a$ satifies
$\frac{1}{d^{3}}<a\leq\frac{1}{d^{2}}$. Moreover $a=\frac{1}{d^{2}}$ if and only if $P_{\alpha}$ are rank one, which gives rise to a SIC-POVM.

Like the mutually unbiased measurements, the authors in \cite{gk14}
explicitly constructed general symmetric informationally complete measurements for arbitrary dimensional spaces. Let $\{F_{\alpha}\}_{\alpha=1}^{d^2-1}$ be a set of $(d^{2}-1)$ Hermitian,
traceless operators on $\mathbb{C}^{d}$, satisfying $\mathrm{Tr}(F_{\alpha}F_{\beta})=\delta_{\alpha,\beta}$, $\alpha,\beta=1,2,\cdots,d^{2}-1$.
Then $d^{2}$ operators
\begin{eqnarray*}
  P_{\alpha} &=& \left\{
  \begin{array}{ll}
\frac{1}{d^{2}}I+t[F-d(d+1)F_{\alpha}] , & \hbox{$\alpha=1,2,\cdots,d^{2}-1,$} \\
\frac{1}{d^{2}}I+t(d+1)F, & \hbox{$\alpha=d^{2},$}
 \end{array}
\right.
\end{eqnarray*}
form a general SIC-POVM measurement, where $F=\Sigma_{\alpha=1}^{d^{2}-1}F_{\alpha}$, $t$ should be chosen such that $P_{\alpha}\geq0$. Corresponding to the construction of GSIC-POVMs, the parameter $a$ is given by
$$
a=\frac{1}{d^{3}}+t^{2}(d-1)(d+1)^{3}.
$$

The entanglement detection based SIC-POVMs has been briefly discussed in Ref.\cite{ra13}, but the method is subject to the existence of SIC-POVMs. However these general symmetric informationally complete measurements do exist for arbitrary dimension $d$ and have many useful applications in quantum information theory. In Ref. \cite{ra14}, based on the calculation of the so-called index of the coincidence, the author derived a number of uncertainty relation inequalities by general SIC-POVMs measurements and  given some SIC-POVM $\mathcal{P}=\{P_{j}\}$ on $\mathbb{C}^{d}$ and density matrix $\rho$, the author \cite{ra14} calcute  the called index of the coincidence $\mathcal{C}(\mathcal{P}|\rho)$, that is,
\begin{equation}
\mathcal{C}(\mathcal{P}|\rho)=\Sigma_{j=1}^{d^{2}}[\mathrm{Tr}(P_{j}\rho)]^{2}
=\frac{(ad^{3}-1)\mathrm{Tr}(\rho^{2})+d(1-ad)}{d(d^{2}-1)}
\end{equation}
Here $\mathcal{C}(\mathcal{P}|\rho)=\frac{ad^{2}+1}{d(d+1)}$ when  $\rho$ is pure.

\medskip
\noindent{\bf 3  Main results and their proofs }

\medskip

\medskip
\noindent{\bf Case $1$ $\mathbb{C}^{d_{1}}\otimes\mathbb{C}^{d_{2}}.$ }

\medskip

\begin{theorem}
Suppose $\rho$ is a density matrix in $\mathbb{C}^{d_{1}}\otimes\mathbb{C}^{d_{2}}$. Let
$\mathcal{P}=\{P_{j}\}_{j=1}^{d_{1}^{2}}$  and
$\mathcal{Q}=\{Q_{k}\}_{k=1}^{d_{2}^{2}}$ be any two sets of  GSIC-POVMs on
$\mathbb{C}^{d_{1}}$ and $\mathbb{C}^{d_{2}}$ with parameters $a_{1},a_{2}$, respectively. Define
$$
J_{1}(\rho,(\mathcal{P},\mathcal{Q}))=\max_{\begin{array}{c}
 \scriptscriptstyle\{P_{j}\}\subseteq \mathcal{P}\\
 \scriptscriptstyle\{Q_{n_{j}}\}\subseteq \mathcal{Q} \end{array}}\Sigma_{j=1}^{d^{2}}\mathrm{Tr}[(P_{j}\otimes Q_{n_{j}})\rho]
$$ \\
Where $d=\min\{d_{1},d_{2}\}$ and for $n_{j}$, there exists a permutation $\sigma\in S_{d_{2}^{2}}$ satifying $\sigma(j)=n_{j}$. If $\rho$ is separable, then $$
J_{1}(\rho)\leq\frac{1}{2}[\frac{a_{1}d_{1}^{2}+1}{d_{1}(d_{1}+1)}+\frac{a_{2}d_{2}^{2}+1}{d_{2}(d_{2}+1)}]
$$.
\end{theorem}

 [{\sf Proof}]. Assume that $\rho=\Sigma_{k=1}^{r}\lambda_{k}|\phi_{k}^{1}\rangle\langle\phi_{k}^{1}|\otimes|\phi_{k}^{2}\rangle\langle\phi_{k}^{2}|
 , \Sigma_{k=1}^{r}\lambda_{k}=1$, $\tilde{J}_{1}(\rho,(\mathcal{P},\mathcal{Q}))=\Sigma_{j=1}^{d^{2}}\mathrm{Tr}[(P_{j}\otimes Q_{n_{j}})\rho]$, $P_{|\phi_{k}^{1}\rangle}(j)=\mathrm{Tr}(P_{j}|\phi_{k}^{1}\rangle\langle\phi_{k}^{1}|)$
\begin{eqnarray*}
&&\tilde{J}_{1}(\rho,(\mathcal{P},\mathcal{Q}))\\
&=&\Sigma_{j=1}^{d^{2}}\mathrm{Tr}[(P_{j}\otimes Q_{n_{j}})\rho]\\
&=& \Sigma_{j=1}^{d^{2}}\Sigma_{k=1}^{r}\lambda_{k}\mathrm{Tr}(P_{j}|\phi_{k}^{1}\rangle\langle\phi_{k}^{1}|)\mathrm{Tr}(Q_{n_{j}}|\phi_{k}^{2}\rangle\langle\phi_{k}^{2}|)\\
&=&
\Sigma_{j=1}^{d^{2}}\Sigma_{k=1}^{r}\lambda_{k}P_{|\phi_{k}^{1}\rangle}(j)Q_{|\phi_{k}^{2}\rangle}(n_{j})\\
&=&
\lambda_{1}\Sigma_{j=1}^{d^{2}}P_{|\phi_{1}^{1}\rangle}(j)Q_{|\phi_{1}^{2}\rangle}(n_{j})+\lambda_{2}\Sigma_{j=1}^{d^{2}}P_{|\phi_{2}^{1}\rangle}(j)Q_{|\phi_{2}^{2}\rangle}(n_{j})+\cdots +\lambda_{r}\Sigma_{j=1}^{d^{2}}P_{|\phi_{r}^{1}\rangle}(j)Q_{|\phi_{r}^{2}\rangle}(n_{j})\\\\
&&\Sigma_{j=1}^{d^{2}}P_{|\phi_{k}^{1}\rangle}(j)Q_{|\phi_{k}^{2}\rangle}(n_{j})\\
&\leq&
\Sigma_{j=1}^{d^{2}}\frac{P_{|\phi_{k}^{1}\rangle}^{2}(j)+Q_{|\phi_{k}^{2}\rangle}^{2}(n_{j})}{2}\\
&\leq&
\frac{\Sigma_{j=1}^{d_{1}^{2}}P_{|\phi_{k}^{1}\rangle}^{2}(j)+\Sigma_{j=1}^{d_{2}^{2}}Q_{|\phi_{k}^{2}\rangle}^{2}(n_{j})}{2}\\
&=&
\frac{1}{2}[\frac{a_{1}d_{1}^{2}+1}{d_{1}(d_{1}+1)}+\frac{a_{2}d_{2}^{2}+1}{d_{2}(d_{2}+1)}]
\end{eqnarray*}

Then $J_{1}(\rho,(\mathcal{P},\mathcal{Q}))=\max\tilde{J}_{1}(\rho,(\mathcal{P},\mathcal{Q}))\leq\frac{1}{2}[\frac{a_{1}d_{1}^{2}+1}{d_{1}(d_{1}+1)}+\frac{a_{2}d_{2}^{2}+1}{d_{2}(d_{2}+1)}]
.\Box$\\
By using the Cauchy-Schwarz inequality, we can obtain stronger bound than in Theorem $1$.

\begin{theorem}
Suppose $\rho$ is a density matrix in $\mathbb{C}^{d_{1}}\otimes\mathbb{C}^{d_{2}}$. Let
$\mathcal{P}=\{P_{j}\}_{j=1}^{d_{1}^{2}}$  and
$\mathcal{Q}=\{Q_{k}\}_{k=1}^{d_{2}^{2}}$ be any two sets of general SIC-POVMs on
$\mathbb{C}^{d_{1}}$ and $\mathbb{C}^{d_{2}}$ with parameters $a_{1},a_{2}$, respectively. Define
$$
J_{2}(\rho,(\mathcal{P},\mathcal{Q}))=\max \tilde{J}_{2}(\rho,(\mathcal{P},\mathcal{Q}))
=\max_{\begin{array}{c}
 \scriptscriptstyle\{P_{j}\}\subseteq \mathcal{P}\\
 \scriptscriptstyle\{Q_{n_{j}}\}\subseteq \mathcal{Q}\} \end{array}}|\Sigma_{j=1}^{d^{2}}\mathrm{Tr}[(P_{j}\otimes Q_{n_{j}})(\rho-\rho^{A}\otimes\rho^{B})]|.
$$ \\
Where $d=\min\{d_{1},d_{2}\}$, $\rho^{A}$($\rho^{B}$) is the reduced density matrix of the first(second) subsystems.
If $\rho$ is separable, then we can

$$J_{2}(\rho)\leq
\sqrt{\frac{a_{1}d_{1}^{2}+1}{d_{1}(d_{1}+1)}-\Sigma_{j=1}^{d_{1}^{2}}[\mathrm{Tr}(P_{j}\rho^{A})]^{2}}
\sqrt{\frac{a_{2}d_{2}^{2}+1}{d_{2}(d_{2}+1)}-\Sigma_{j=1}^{d_{2}^{2}}[\mathrm{Tr}(Q_{j}\rho^{B})]^{2}}.$$

\end{theorem}

[{\sf Proof}]. Assume that $\rho=\Sigma_{k=1}^{r}p_{k}\rho_{k}^{A}\otimes\rho_{k}^{B}$, $0\leq p_{k}\leq1, \Sigma_{k=1}^{r}p_{k}=1$, where $\rho_{k}^{A}$ and $\rho_{k}^{B}$ are the pure density matrix acting on the first and second subsystem. Thus we can get $\rho^{A}=\Sigma_{k=1}^{r}p_{k}\rho_{k}^{A},\rho^{B}=\Sigma_{k=1}^{r}p_{k}\rho_{k}^{B}$. Let $\tilde{J}_{2}=|\Sigma_{j=1}^{d^{2}}\mathrm{Tr}[(P_{j}\otimes Q_{n_{j}})(\rho-\rho^{A}\otimes\rho^{B})]|$, $p_{s,t}=\sqrt{p_{s}p_{t}}$.
Then
\begin{eqnarray*}
&&\Sigma_{j=1}^{d^{2}}|\mathrm{Tr}[(P_{n_{j}}\otimes Q_{n_{j}})(\rho-\rho^{A}\otimes\rho^{B})]|\\
&=&\Sigma_{j=1}^{d^{2}}|\mathrm{Tr}[(P_{n_{j}}\otimes Q_{n_{j}})(\frac{1}{2}\Sigma_{s,t=1}^{r}p_{s}p_{t}(\rho_{s}^{A}-\rho_{t}^{A})\otimes(\rho_{s}^{B}-\rho_{t}^{B}))]|\\
&\leq& \Sigma_{j=1}^{d^{2}}\Sigma_{s,t=1}^{r}\frac{1}{2}|\mathrm{Tr}[\sqrt{p_{s}p_{t}}P_{n_{j}}(\rho_{s}^{A}-\rho_{t}^{A})]||\mathrm{Tr}[\sqrt{p_{s}p_{t}}Q_{n_{j}}(\rho_{s}^{B}-\rho_{t}^{B})]|\\
&\leq&
\sqrt{\frac{\Sigma_{j,s,t}\{\mathrm{Tr}[p_{s,t}P_{n_{j}}(\rho_{s}^{A}-\rho_{t}^{A})]\}^{2}}{2}}\sqrt{\frac{\Sigma_{j,s,t}\{\mathrm{Tr}[p_{s,t}Q_{n_{j}}(\rho_{s}^{B}-\rho_{t}^{B})]\}^{2}}{2}}.
\end{eqnarray*}
So we can get
\begin{eqnarray*}
&&J_{2}(\rho)=\max \tilde{J}_{2}\\
&\leq&
\Sigma_{j=1}^{d^{2}}|\mathrm{Tr}[(P_{n_{j}}\otimes Q_{n_{j}})(\rho-\rho^{A}\otimes\rho^{B})]|\\
& \leq &
\sqrt{\frac{a_{1}d_{1}^{2}+1}{d_{1}(d_{1}+1)}-\Sigma_{j=1}^{d_{1}^{2}}[\mathrm{Tr}(P_{j}\rho^{A})]^{2}}
\sqrt{\frac{a_{2}d_{2}^{2}+1}{d_{2}(d_{2}+1)}-\Sigma_{j=1}^{d_{2}^{2}}[\mathrm{Tr}(Q_{j}\rho^{B})]^{2}}.\Box
\end{eqnarray*}

In order to formulate the validity and power of entanglement detection, we consider the examples in the following.

Example 1.  Let us consider the isotropic states, which are locally unitarily equivalent to a maximally entangled state mixed with white noise:



 $$\rho=\rho_{iso}=\alpha|\phi^{+}\rangle\langle\phi^{+}|+\frac{1-\alpha}{d^{2}}\mathds{I},$$
where $ 0\leq\alpha\leq1, |\phi^{+}\rangle=\frac{\Sigma_{i=1}^{d}|ii\rangle}{d}$,
$$J_{1}(\rho,(\mathcal{P},\mathcal{Q}))=
\tilde{J}_{1}(\rho_{iso},(\mathcal{P},\mathcal{\overline{P}}))=\Sigma_{j=1}^{d^{2}}\mathrm{Tr}[(P_{j}\otimes \overline{P}_{j})\rho]=\alpha da+\frac{1-\alpha}{d^{2}}=(da-\frac{1}{d^{2}})\alpha+\frac{1}{d^{2}},
$$
where $\tilde{J}_{1}(\rho_{iso},(\mathcal{P},\mathcal{\overline{P}}))$ is montone increasing in $\alpha$. If $\alpha>\frac{1}{d+1}$, $\tilde{J}_{1}(\rho_{iso},(\mathcal{P},\mathcal{\overline{P}}))> \frac{ad^{2}}{d(d+1)}$ and $\rho_{iso}$ must be entangled by our theorem. Then Theorem 1 can detect all the entanglement of the isotropic states, because it has been proven $\rho_{iso}$ is entangled for $\alpha>\frac{1}{d+1}$, and separable for $\alpha\leq\frac{1}{d+1}$.\cite{BK}
So we can get that for $\rho_{iso}$, Theorem $1$ gives a necessary and sufficient separable criterion.

Example 2. Next consider the $d$- dimensional Bell-diagonal states
$$\rho_{Bell}= \Sigma_{s,t=0}^{d-1}c_{st}|\phi_{st}\rangle\langle\phi_{st}|,$$

where$c_{st}\geq 0,\Sigma_{s,t=0}^{d-1}c_{st}=1,
|\phi_{st}\rangle= (U_{st}\otimes\mathbb{I})|\phi^{+}\rangle, U_{st}= \Sigma_{j=0}^{d-1}\sigma_{d}^{sj}|j\rangle\langle\ j\oplus t|, \sigma_{d}= e^{\frac{2\Pi\sqrt[2]{-1}}{d}},$

and $\oplus$ denoting  $(j+t)$ mod $d$.

By some simple calculations, there are
$$
\rho_{Bell}^{A}=\rho_{Bell}^{B}
=\frac{\mathbb{I}}{d}.
$$

Assume $c_{0}=\min\{c_{st}\}, c_{d^{2}}=\max\{c_{st}\}.$

 $$
 \tilde{J}_{1}(\rho_{Bell},(\mathcal{P},\mathcal{Q}))=\Sigma_{j=1}^{d^{2}}\mathrm{Tr}[(P_{j}\otimes {Q}_{n_{j}})\rho_{Bell}]\geq \tilde{J}_{1}(c_{d^{2}}|\phi_{c_{d^{2}}}\rangle\langle\phi_{c_{d^{2}}}|,(\mathcal{P},\mathcal{Q})).
 $$

Then
$$
\begin{array}{c}
 J_{1}(\rho_{Bell},(\mathcal{P},\overline{U_{c_{0}}^{+}\mathcal{P}U_{c_{0}}}))\geq J_{1}(|\phi_{c_{0}}\rangle\langle\phi_{c_{0}}|,(\mathcal{P},\overline{U_{c_{0}}^{+}\mathcal{P}U_{c_{0}}}))\\
 J_{1}(\rho_{Bell},(\mathcal{P},\overline{U_{c_{d^{2}}}^{+}\mathcal{P}U_{c_{d^{2}}}}))\geq J_{1}(|\phi_{c_{d^{2}}}\rangle\langle\phi_{c_{d^{2}}}|,(\mathcal{P},\overline{U_{c_{d^{2}}}^{+}\mathcal{P}U_{c_{d^{2}}}}))
 \end{array}
$$

From Theorem 1, we can get $\rho_{Bell}$ is entangled if
$c_{d^{2}}>\frac{{ad^{2}+1}}{d^2(d+1)a}.$

From Theorem 2, we can get $\rho_{Bell}$ is entangled if  $0\leq c_{0} \leq\frac{1}{d^{2}},$ or $\frac{1}{d^{2}}\leq c_{d^{2}}\leq\frac{d-d^{3}a+2}{d^{3}a(d+1)}$, or $\frac{1}{ad^{3}}\leq c_{d^{2}}\leq1$.

It is obvious that the values of $a$ affect the performance of the entanglement detection of Theorem 1 and Theorem 2. When  $a$ gets larger or smaller, the criteria can detect more entanglement.

\begin{remark}

Here we will talk something about these upper bounders. For Theorem 1 and Theorem 2, if $$d_{1}=d_{2}=d, \rho=\rho_{iso},
(\mathcal{P},\mathcal{Q})=(\mathcal{P},\mathcal{\overline{P}}).
$$

When $\alpha=\frac{1}{d+1}$, $\tilde{J}_{1}(\rho_{iso},(\mathcal{P},\mathcal{\overline{P}}))= \frac{ad^{2}+1}{d(d+1)}, \tilde{J}_{2}(\rho_{iso}, (\mathcal{P},\mathcal{\overline{P}}))= \frac{ad^{2}+1}{d(d+1)}-\frac{1}{d^{2}}.$

If $d_{1}\neq d_{2}, d_{1}<d_{2}$ and the bound we get in the theorem 1 can  be reached, then

$$d_{2}^{2}+d_{2}-2d_{1}^{2}<0, a_{2} \in(\frac{d_{1}^{2}+d_{2}}{d_{2}^{2}(d_{2}^{2}-d_{1}^{2})},\frac{1}{d_{2}^{2}}), \mathrm{Tr}(\rho^{A})=\mathrm{Tr}(\rho^{B}).
$$
For example, the bounder can not be reached in $\mathbb{C}^{2}\otimes\mathbb{C}^{n_{1}}$ and $\mathbb{C}^{3}\otimes\mathbb{C}^{n_{2}},$ where $n_{1}\geq3, n_{2}\geq4$. So it is also an interesting problem to construct other new and efficient separable criteria.
\end{remark}

\begin{remark}
In some case, our result is more effective than the results in \cite{clf15}.
 Consider
 $$
 \rho= \frac{\mathbb{I}\otimes\mathbb{I}}{4}+\frac{1}{16}\sigma_{1}\otimes\sigma_{1}-\frac{1}{4}\sigma_{2}\otimes\sigma_{2}+\frac{1}{16}\sigma_{3}\otimes\sigma_{3}.
 $$

 And a class
of measurements $(\mathcal{P},\mathcal{P})$, satifying
$$
\mathcal{P}=\{\frac{\mathbb{I}}{4}+t(F-6F_{1}),\frac{\mathbb{I}}{4}+t(F-6F_{2}),\frac{\mathbb{I}}{4}+t(F-6F_{3}), \frac{\mathbb{I}}{4}+3tF \},
$$

$$
F=\Sigma F_{i}, F_{i}=\frac{\sigma_{i}}{\sqrt[2]{2}}
$$
and
$\sigma_{i}^{,}s$
are three pauli matrix,
$i=1,2,3.$  Here $\frac{ad^{2}+1}{d(d+1)}=\frac{1}{4}+18t^{2}$ and $t\in(-\frac{1}{12}\sqrt[2]{\frac{2}{3}}, \frac{1}{12}\sqrt[2]{\frac{2}{3}}).$

The criterion in  Ref. \cite{clf15} can not be used to detect $\rho$, but our Theorem 1 can detect the entanglement of $\rho.$ In fact, for the pair
of measurements $(\mathcal{P},\mathcal{P})$, using the criterion in  Ref. \cite{clf15}, we can get

$$
\tilde{J}(\rho,(\mathcal{P},\mathcal{P}))=\Sigma_{j=1}^{4}\mathrm{Tr}[(P_{j}\otimes p_{j})\rho]\\
=\frac{1}{4}-9t^{2}<\frac{1}{4}+18t^{2}
$$
which can not decide the entanglement of $\rho.$

But by Theorem 1, the state $\rho$ is entangled because
\begin{eqnarray*}
&&\tilde{J}(\rho,(\mathcal{P},\mathcal{P}))\\
&=&\Sigma_{j=1}^{4}\mathrm{Tr}[(P_{j}\otimes p_{n_{j}})\rho]\\
&=&
\mathrm{Tr}[(P_{1}\otimes p_{1})\rho]+\mathrm{Tr}[(P_{2}\otimes p_{4})\rho]+\mathrm{Tr}[(P_{3}\otimes p_{3})\rho]+\mathrm{Tr}[(P_{4}\otimes p_{2})\rho]\\
&=&
\frac{1}{4}+22t^{2}>\frac{1}{4}+18t^{2}
\end{eqnarray*}
\end{remark}

\medskip
\noindent{\bf Case$2$ $\mathbb{C}^{d_{1}}\otimes\mathbb{C}^{d_{2}}\cdots\otimes\mathbb{C}^{d_{n}}.$}

\medskip
For multipartite systems that the definition of entanglement is not unique. So we discuss it with the the notions of $k$-partite entanglement or $k$-nonseparability for a given partition and unfixed partition,respectively \cite{s1,g09}. A pure state $\mathbf{|\phi\rangle}$ of a $n$-partite system is called $k$-separable if it can be written as a tensor product of $k$ vectors, i.e. $\mathbf{|\phi\rangle}=|\phi\rangle_{1}\otimes|\phi\rangle_{2}\otimes\cdots\otimes|\phi\rangle_{k}$.
The states which do not contain any entanglement are called fully separable. In addition, those states whose entanglement ranges over all $n$ parties are called genuine multipartite entangled states. The generalization to mixed states is direct: A mixed state is called $k$-separable if it can be written as a convex combination of $k$-separable states $\rho=\Sigma_{k=1}^{r}p_{k}\rho_{k}$, where $\rho_{k}$ are $k$-separable pure states. In the following, we have two criteria for multipartite systems of different dimensions and also argue $k$-nonseparability for a given partition of $n$-partite system.

\begin{theorem}
Suppose $\rho$ is a density matrix in $\mathbb{C}^{d_{1}}\otimes\mathbb{C}^{d_{2}}\otimes\cdots\otimes\mathbb{C}^{d_{n}}$,
and $\{\mathcal{P}^{(i)}\}$ are $n$ sets of GSIC-POVMs in
$\mathbb{C}^{d_{i}}$ with parameters $a_{i},i=1,2,\cdots,n$,
where$\{\mathcal{P}^{(i)}\}=\{P_{j}^{(i)}\}_{j=1}^{d_{i}^{2}}$. Define

$$
J_{3}(\rho)=\max_{\{P_{n_{j}}^{(i)}\}\subseteq\{\mathcal{P}_{j}^{(i)}\}}\Sigma_{j=1}^{d}\mathrm{Tr}(\otimes_{i=1}^{n}P_{n_{j}}^{(i)}\rho). $$
Here $d=\min\{d_{1}^{2},d_{2}^{2},\cdots,d_{n}^{2}\}$. If $\rho$ is fully separable, then

$$
J_{3}(\rho)\leq\frac{1}{n}\Sigma_{i=1}^{n}[\frac{a_{i}d_{i}^{2}+1}{d_{i}(d_{i}+1)}]
$$
\end{theorem}

[{\sf Proof}]. Let $\rho=\Sigma _{k=1}^{r}p_{k}P_{k}$, with $\Sigma_{k=1}^{r}p_{k}=1$, be a fully separable density matrix, where $\rho_{k}=\otimes_{i=1}^{n}|\phi_{ik}\rangle\langle\phi_{ik}|$. Since

\begin{eqnarray*}
\Sigma_{j=1}^{d}\mathrm{Tr}[(\otimes_{i=1}^{n}P_{n_{j}}^{(i)})\rho_{k}]
& = & \Sigma_{j=1}^{d}\mathrm{Tr}[(\otimes_{i=1}^{n}P_{n_{j}}^{(i)})(\otimes_{i=1}^{n}|\phi_{ik}\rangle\langle\phi_{ik}|)]\\
& = & \Sigma_{j=1}^{d}[\Pi_{i=1}^{n}\mathrm{Tr}(P_{n_{j}}^{(i)}|\phi_{ik}\rangle\langle\phi_{ik}|)]\\
& \leq &
\Sigma_{i=1}^{n}\Sigma_{j=1}^{d}\frac{[\mathrm{Tr}(P_{n_{j}}^{(i)}|\phi_{ik}\rangle\langle\phi_{ik}|)]^{2}}{n}
\end{eqnarray*}
where we use the inequality in \cite{s4}
$$x_{1}x_{2}\cdots x_{n}\leq[\frac{\Sigma_{i=1}^{n}(x_{i})^{2}}{n}]^{\frac{n}{2}},x_{i}\geq0,i=1,2,\cdots n $$

Then through th equality$(1)$, we can get $J_{3}(\rho)\leq\frac{1}{n}\Sigma_{i=1}^{n}[\frac{a_{i}d_{i}^{2}+1}{d_{i}(d_{i}+1)}].\Box$

\begin{theorem}
Assume that $\rho$ is a density matrix in $\mathbb{C}^{d_{1}}\otimes\mathbb{C}^{d_{2}}\otimes\cdots\otimes\mathbb{C}^{d_{n}}$,
and $\{P_{j}^{(i)}\}_{j=1}^{d_{i}^{2}}$ are $n$ sets of GSIC-POVMs on
$\mathbb{C}^{d_{i}}$ with parameters $a_{i},i=1,2,\cdots,n$.

If $\rho$ is fully separable, then

$$
J_{3}(\rho)\leq \min_{i\neq j}\sqrt{\frac{a_{i}d_{i}^{2}+1}{d_{i}(d_{i}+1)}} \sqrt{\frac{a_{j}d_{j}^{2}+1}{d_{j}(d_{j}+1)}}
$$
\end{theorem}
[{\sf Proof}].  Let $\rho=\Sigma _{k=1}^{r}p_{k}P_{k}$ be a fully separable pure state, where $\Sigma_{k=1}^{r}p_{k}=1$.

\begin{eqnarray*}
I(\rho)  & = &
\Sigma_{j=1}^{d} \Sigma_{k=1}^{r} p_{k} \mathrm{Tr}[(\otimes_{i=1}^{n}P_{n_{j}}^{(i)})P_{k}]\\
& = &
\Sigma_{k=1}^{r}\Sigma_{j=1}^{d} p_{k} \mathrm{Tr}[(\otimes_{i=1}^{n}P_{n_{j}}^{(i)})(\otimes_{i=1}^{n}|\phi_{i}\rangle\langle\phi_{i}|)]\\
 &= &
\Sigma_{k=1}^{r} \Sigma_{j=1}^{d} p_k \Pi_{i=1}^{n} \mathrm{Tr}(P_{n_{j}}^{(i)}|\phi_{i}\rangle\langle\phi_{i}|)
\end{eqnarray*}

Then using the Cauchy-Schwarz inequality, we can get
$$
I(\rho)\leq \sqrt{\Sigma_{j=1}^{d}[\mathrm{Tr}(P_{n_{j}}^{(i)}|\phi_{i}\rangle\langle\phi_{i}|)]^2}
\sqrt{\Sigma_{j=1}^{d}[\mathrm{Tr}(P_{n_{j}}^{(i')}|\phi_{i'}\rangle\langle\phi_{i'}|)]^2},
$$
where $i\neq i'$ and using the equality $(1)$, we finally get
$$
J_{3}(\rho)= \max I(\rho)\leq \min_{i\neq j}\sqrt{\frac{a_{i}d_{i}^{2}+1}{d_{i}(d_{i}+1)}} \sqrt{\frac{a_{j}d_{j}^{2}+1}{d_{j}(d_{j}+1)}}.\Box
$$

In partically, the criterion in Ref. \cite{clf15} is the special case of Theorem $3$. What's more, we can use Theorem $3$ and Theorem $4$ straightforward to detect $k$-nonseparable states with respect to a fixed partition. For an $n$-partite state $\rho$ in $\mathbb{C}^{m_{1}}\otimes\mathbb{C}^{m_{2}}\otimes\cdots\otimes\mathbb{C}^{m_{n}}=\mathbb{C}^{d_{1}}\otimes\mathbb{C}^{d_{2}}\otimes\cdots\otimes\mathbb{C}^{d_{k}}$, if there are  $k$ sets of GSIC-POVMs $\{\mathcal{P}^{(i)}\}$  on
$\mathbb{C}^{d_{i}}$ with parameters $a_{i}$ such that
$$
\Sigma_{j=1}^{d}\mathrm{Tr}(\otimes_{i=1}^{k}P_{n_{j}}^{(i)}\rho)>\frac{1}{k}\Sigma_{i=1}^{k}\frac{a_{i}d_{i}^{2}+1}{d_{i}(d_{i}+1)}
$$

or

$$
\Sigma_{j=1}^{d}\mathrm{Tr}(\otimes_{i=1}^{k}P_{n_{j}}^{(i)}\rho)>
\min_{1\leq i\neq j\leq k} \sqrt{\frac{a_{i}d_{i}^{2}+1}{d_{i}(d_{i}+1)}}\sqrt{\frac{a_{j}d_{j}^{2}+1}{d_{j}(d_{j}+1)}} $$
\\
for some $\{P_{n_{j}}^{(i)}\}_{j=1}^{d^2}\subseteq\{\mathcal{P}^{(i)}\}$, then $\rho$ is $k$-nonseparable in $\mathbb{C}^{d_{1}}\bigotimes\mathbb{C}^{d_{2}}\otimes\cdots\otimes\mathbb{C}^{d_{k}}$, where $d=min\{d_{1}^{2},d_{2}^{2},\cdots,d_{k}^{2}\}$ and $i=1,2,\cdots,k$.

\medskip
\noindent{\bf 4  Conclusions and discussions}

\medskip

In summary, we have analyzed the separability problem based on the GSIC-POVMs and presented separability criteria in $\mathbb{C}^{d_{1}}\otimes\mathbb{C}^{d_{2}}$ and $\mathbb{C}^{d_{1}}\otimes\mathbb{C}^{d_{2}}\cdots\otimes\mathbb{C}^{d_{n}}$.
Our results are useful. First, our criteria are suitable for arbitrary dimension $d$ because the GSIC-POVMs do exist for arbitrary dimension $d$. Second, The criteria in this paper could detect the separability of arbitrary high dimensional bipartite systems and multipartite systems of different dimensions only by less joint local measurements to reduce the complexity of experimental implementation, and our result is more effective than the results in \cite{clf15} in Remark 2. It would be interesting to construct efficient criteria for entanglement detection using quantum measurements.

\bigskip
\noindent{\bf Acknowledgments}\, \,
 This work is supported by the NSFC through Grants No.11475178, No.11571119 and No.11135006. And we would like to thank Prof. Shao-Ming Fei at School of Mathematical Sciences, Capital Normal University for helpful discussion.


\begin{thebibliography}{99}
\bibitem{s1} R. Horodecki, P. Horodecki, M. Horodecki, and K. Horodecki, Rev. Mod. Phys. \textbf{81}, 865 (2009).

\bibitem{g09} O. Guhne, G. Toth, Phys. Rep. \textbf{474}, 1 (2009).

\bibitem{b64} J.S. Bell, Physics(N.Y.) \textbf{1}, 195 (1964).

\bibitem{p96} A. Peres, Phys. Rev. Lett. \textbf{77}, 1413 (1996); M. Horodecki, P. Horodecki, and R. Horodecki, Phys. Lett. A \textbf{223}, 1 (1996).
\bibitem{r1} O. Rudolph, Phys. Rev. A \textbf{67}, 032312 (2003).
\bibitem{r2} K. Chen and L. A. Wu, Phys. Lett. A \textbf{306}, 14 (2002).

\bibitem{r3} S. Albeverio, K. Chen, and S. M. Fei, Phys. Rev. A \textbf{68}, 062313 (2003).

\bibitem{BK} R. A. Bertlmann, K. Durstberger, B. C. Hiesmayr, and P. Krammer, Phys. Rev. A \textbf{72}, 052331 (2005).
\bibitem{c1} O. Guhne, P. Hyllus, O. Gittsovich, and J. Eisert, Phys. Rev. Lett. \textbf{99}, 130504 (2007).

\bibitem{c2} J. D. Vicente, Quant. Inf. Comput. \textbf{7}, 624 (2007).

\bibitem{t00} B. Terhal, Phys. Lett. A \textbf{271}, 319 (2000); M. Lewenstein, B. Kraus, J.I. Cirac and P. Horodecki, Phys. Rev. A \textbf{62}, 052310 (2000).

\bibitem{Bell} N. Gisin, Phys. Lett. A \textbf{154}, 201 (1991); M.J. Zhao, T. Ma,  S.M. Fei and Z.X. Wang, Phys. Rev. A \textbf{83}, 052120 (2011).

\bibitem{spe12} C. Spengler, M. Huber, S. Brierley, T. Adaktylos, and B. C. Hiesmayr, Phys. Rev. A \textbf{86}, 022311 (2012).

\bibitem{s2} W. K. Wootters and B. D. Fields, Ann. Phys. (N.Y.) \textbf{191}, 363 (1989).

\bibitem{s3} B. Chen, T. Ma, and S.M. Fei, Phys. Rev. A \textbf{89}, 064302 (2014).

\bibitem{ka14} A. Kalev and G. Gour, New J.Phys. \textbf{16}, 053038 (2014).

\bibitem{ra15} A. E. Rastegin, Open Sys. Inf. Dyn. \textbf{22}, 1550005(2015).

\bibitem{s4} L. Liu, T. Gao, and F. L. Yan, arXiv: 1501.01717[quant-ph] (2015).

\bibitem{s5} J. M. Renes, R. Blume-Kohout, A. J. Scott, C. M. Caves, J. Math. Phys. \textbf{45}, 2171 (2004); A. J. Scott, M. Grassl, J. Math. Phys. \textbf{51}, 042203 (2010).

\bibitem{s7} D. M. Appleby, Optics and Spectroscopy \textbf{103},416 (2007).

\bibitem{gk14} G. Gour and A. Kalev, J. Phys. A: Math. Theor \textbf{47}, 335302 (2014).

\bibitem{clf15} B. Chen, Tao. Li, and S.M. Fei, Quantum. information. processing \textbf{14}, 2281-2290 (2015).

\bibitem{shen1} Shu-Qian Shen, Ming Li, and Xue-Feng Duan, Phys. Rev. A \textbf{91}, 012326 (2015).

\bibitem{ra13} A. E. Rastegin, Eur. Phys. J. D \textbf{67}, 269 (2013).

\bibitem{ra14} A. E. Rastegin, Phys. Scr. \textbf{89}, 085101 (2014).




\end{thebibliography}
\end{document}